\documentclass[prl,superscriptaddress,showpacs,twocolumn]{revtex4}
\usepackage{graphicx}
\def\figwidth{8cm}
\def\smallwidth{7cm}
\begin{document}
\title{Variable range hopping and quantum creep in one dimension}

\author{Thomas Nattermann}
\affiliation{Institut f\"ur Theoretische Physik, Universit\"at zu
K\"oln, Z\"ulpicher Str. 77 D-50937 K\"oln, Germany}

\author{Thierry Giamarchi}
\affiliation{Universite de Geneve, DPMC, 24 Quai Ernest Ansermet,
CH-1211 Geneve 4, Switzerland}

\author{Pierre Le Doussal}
\affiliation{CNRS-Laboratoire de Physique Theorique de l'Ecole Normale Superieure,
24 Rue Lhomond, Paris 75231 France.}
\date{\today}

\begin{abstract}
We study the quantum non linear response to an applied electric
field $E$ of a one dimensional pinned charge density wave or
Luttinger liquid in presence of disorder. From an explicit
construction of low lying metastable states and of bounce
instanton solutions between them, we demonstrate quantum creep $v
= e^{- c/E^{1/2}}$ as well as a sharp crossover at $E=E^*$ towards
a linear response form consistent with variable range hopping
arguments, but dependent only on electronic degrees of freedom.
\end{abstract}
\maketitle

Computing the response of a disordered elastic system to an
external driving force is a long standing problem. This is of
theoretical importance and also relevant for a host of
experimental systems, both classical and quantum. For classical
systems, typical experimental realizations are domain walls
\cite{lemerle_domainwall_creep,tybell_ferro_creep} and vortex
lattice in type II superconductors
\cite{blatter_vortex_review,nattermann_vortex_review}. Pinned
quantum crystals are charge or spin density waves
\cite{gruner_revue_cdw}, Wigner crystal in two dimensional
electron gas \cite{andrei_wigner_2d,willett_wigner_resistivity}
and disordered Luttinger liquids \cite{giamarchi_quantum_revue}.
In the absence of quantum or thermal fluctuations disorder leads
to pinning or localization. It was initially believed that thermal
activation over barriers between pinned states would result
\cite{anderson_kim} in $v(F) \sim \sigma F$ albeit with an
exponentially small mobility $\sigma$. However, the glassy nature
of such disordered elastic systems leads instead to divergent
barriers and to a non linear response
\cite{ioffe_creep,nattermann_rfield_rbond} of the form $v =
\text{exp}(- \beta F^{- \mu})$ known as creep
\cite{blatter_vortex_review}.

In quantum disordered systems barriers between the many metastable
states can be overcome by thermal and quantum activation.
Determination of the relation $v(F)$ is thus an even more
difficult and mostly open question. Two main issues arise: (i)
does one recovers a quantum creep formula at $T=0$ when the system
can unpin via quantum tunnelling over barriers; (ii) does one
recovers linear response at $T>0$, $v(F)/F \to \sigma$ and what is
the $T$ dependence of the conductivity $\sigma$. Although these
questions have been answered in details via controlled instanton
calculations for pure systems such as the Sine-Gordon model
\cite{maki_instanton_periodic_long,hida_tunnel_sinegordon,hida_tunnel_sinegordon_finiteT},
with and without dissipation, no controlled method has been found
for the disordered problem. Results were obtained using physical
arguments for very disordered electronic systems
\cite{shklovskii_conductivity_coulomb}. The renormalization method
used for creep in classical systems \cite{chauve_creep_long} was
extended to quantum problems \cite{gorokhov_quantum_creep}, but
suffers from the same limitations \cite{balents_frg_finiteT}. The
conductivity of charge density waves was studied by Larkin and Lee
\cite{larkin_tunnelling_1d}, but only in a strong pinning regime
considering tunnelling around single impurities.

In this paper we study the driven quantum dynamics of a pinned 1D
charge density wave or of 1D interacting electrons (Luttinger
liquid (LL)) in the localized phase, performing a controlled
calculation of the tunnelling rates. It is known that this system
renormalizes to strong disorder where the (classical) ground state
can be found exactly and low lying kink-like excited states
constructed. We then study instanton (bounce) solutions and
estimate the semiclassical tunnelling rate between these states,
in presence of an applied (electric) field. This demonstrates a
quantum creep law $v = e^{- c/E^{1/2}}$ at zero temperature. At
small non zero temperature we show that a sharp crossover occurs
between quantum creep for $E>E^*$ and linear response for $E<E^*$.
The temperature dependence of the conductivity is of the form
$\sigma \propto e^{-c/T^{1/2}}$ consistent with Mott's variable
range hopping (VRH) arguments \cite{mott_metal_insulator}. Applied
to the Luttinger liquid, this extends in a precise way the
validity of VRH formula to interacting electrons in $d=1$. Note
that here contrarily to standard VRH arguments the prefactor $c$
of the temperature dependence in the exponential is determined by
the electronic degrees of freedom, and is not dependent in an
essential way on coupling to other degrees of freedom such as
phonons. This leads to quite different energy scales for $c$ than
the standard VRH mechanism.

We consider the Hamiltonian of a charge density wave where the
density has a sinusoidal modulation
\begin{equation}
 \rho(x) = \rho_0 \cos(Q x - \phi(x))
\end{equation}
where $\phi$ is the phase of the charge density wave. The phase
$\phi$ obeys the standard phase action \cite{fukuyama_pinning}
\begin{equation} \label{eq:startkin}
 \frac{S}{\hbar} = \int_0^{L} dx \int_0^{\beta\hbar u} dy
 \frac{1}{2\pi K}\left[(\partial_y\phi)^2 + (\partial_x\phi)^2
 \right]
\end{equation}
where $u$ is the velocity, $y= u \tau$ and $\beta$ the inverse of
the temperature. Furthermore the system has a short distance
cutoff (lattice spacing) $\alpha$. The disorder is modelled by a
random potential $V(x)$ coupled to the density by $H = -\int dx
V(x)\rho(x)$. Assuming that $\phi$ varies slowly at the scale
$Q^{-1}$, we can only retain the Fourier components of $V(x)$
close to $Q$ \cite{giamarchi_loc}. This leads to the action
\begin{eqnarray} \label{eq:startdis}
 S_{\text{dis}}/\hbar &=& - \frac12
 \int \frac{dx dy A(x)}{2\pi K \alpha^2} e^{i(\phi(x,y)-\zeta(x))} + \text{h.c.}
\end{eqnarray}
where we represent the disorder with a random amplitude $A$ and
phase $\zeta$, which are both slowly varying variables. For a
Gaussian disorder initially, the disorder $\xi(x) =
A(x)e^{i\zeta(x)}$ obeys
 $\overline{\xi(x)\xi(x')^*} = D \delta(x-x')$
other averages are zero. Adding an external electric field $E$ to
the system adds to the action
 \begin{equation} \label{eq:starte}
 S_{E}/\hbar = \int dx dy \tilde E  \phi(x,y)
\end{equation}
with $\tilde E = E \rho_0/(Q u \hbar)$. The action
(\ref{eq:startkin}-\ref{eq:starte}) also describes a LL in
presence of disorder \cite{giamarchi_loc,giamarchi_quantum_revue}.
In that case $Q = 2\pi\rho_0 = 2 k_F$ where $k_F$ is the Fermi
wavevector for fermions. $K$ is the standard Luttinger parameter
that describe the interactions effects ($K=1$ for noninteracting
electrons and $K<1$ for repulsive interactions). Our study thus
directly gives the conductivity of disordered LLs. In that case
the pinning of the phase variable $\phi$ corresponds to the
Anderson localization of the system.

At $T=0$ the disorder is a relevant variable. It pins the phase
$\phi$. In the ground state the phase $\phi$ varies by a quantity
of order $2\pi$ over a distance $\xi_{\text{loc}}$ which is the
pinning length of the charge density wave \cite{fukuyama_pinning}
or the localization length in the presence of interactions for the
interacting particles
\cite{giamarchi_loc,giamarchi_quantum_revue}. To determine the
dynamics of this model, we renormalize the system up to a point
where the disorder is of order one. Since we are interested in the
limit of very low temperatures and fields, we can renormalize the
action in the absence of $E$ and at $T=0$. The flow in that case
is well known \cite{giamarchi_loc,glatz_cdw_finitetemp} and we do
not reproduce it here. The disorder $D$ scales to strong coupling,
and the parameter $K$ decreases. We stop the flow at the
lengthscale $l^*$ for which $A \sim 1$. At that lengthscale the
disorder being of order one, the pinning length is of the order of
the lattice spacing $\alpha$. The original localization length of
the system is thus $\xi_{\text{loc}} = \alpha e^{l^*}$ The
electric field is also renormalized and becomes
 $\frac{2\epsilon}{K^* \alpha} = \tilde E(l^*) = \tilde E e^{2 l^*}$
and time and space are rescaled by a factor $e^{-l^*}$. In what
follows we denote with a star the renormalized quantities at the
scale $\xi_{\text{loc}}$. Since $u$ also renormalizes one can
absorbs this renormalization by rescaling the time by $u^*/u$
which changes $u \to u^*$ in all the above expressions.

To study the dynamics we consider
(\ref{eq:startkin}-\ref{eq:starte}) with the \emph{renormalized}
parameters. Although we stopped the flow when the disorder is of
order one we assume that we are truly at strong disorder and can
thus consider that the \emph{amplitude} of the disorder is very
large. The main effects thus come from the fluctuations of the
random phase of the disorder. In order to perform a semiclassical
approximation for the dynamics one must first determine the
(classical) ground state of the renormalized system for $E=0$. The
disorder being time independent the action is minimized by
$\partial_\tau \phi =0$. It is convenient to go back to a lattice
description. The energy on the lattice is
\begin{equation}
 \frac{H}{u^* \hbar} = \frac{1}{2\pi K^* \alpha} \sum_{i=1}^{N} \left[(\phi_{i+1}-\phi_{i})^2 -
   A^* \cos(\phi_i - \zeta_i)\right]
\end{equation}
with $N=L/\xi_{\text{loc}}$ and we take $-\pi < \zeta_i \leq \pi$.
Since the renormalized disorder $A^*$ in (\ref{eq:startdis}) can
be considered to be large ($A^*=1$), to minimize the cosine term
one needs to take $\phi_i = \zeta_i + 2\pi n_i$ where $n_i$ are
integers. The energy becomes
\begin{eqnarray}
 \frac{H}{u^* \hbar} &=& \frac{2 \pi}{K^* \alpha} \sum_{i=1}^{N}
 \left(n_{i+1}-n_i - f_i \right)^2  - \frac{N}{2\pi K^* \alpha}
\end{eqnarray}
with $f_i = (\zeta_{i+1}-\zeta_i)/(2 \pi)$, $-1<f_i<1$. Contrarily
to higher dimension, here in $d=1$ there is no frustration and one
can minimize the action for all bonds simultaneously (i.e. all
pairs $\Delta n_i = n_{i+1}-n_i$) by choosing \cite{glatz_diplom}:
\begin{equation}
 n^0_i = m_0 + \sum_{j<i} \left[f_i \right]
\end{equation}
where $m_0$ is an arbitrary integer and $[x]$ denotes the closest
integer to $x$. $[x]=0$ if $-1/2<x<1/2$ and $[x]=1$ (resp.
$[x]=-1$) for $x>1/2$ (resp $x<-1/2$). The values $[f_i]$ thus
completely characterize the ground state. Here one takes the
$\zeta_i$ uniformly distributed, hence the $n^0_i$ perform a
random walk and the ground state has roughness exponent $1/2$
(i.e. $\phi(L)$ scales as $\phi(L) \sim L^{1/2}$) in agreement
with other calculations \cite{giamarchi_columnar_variat}.

\begin{figure}
 \centerline{\includegraphics[width=\smallwidth]{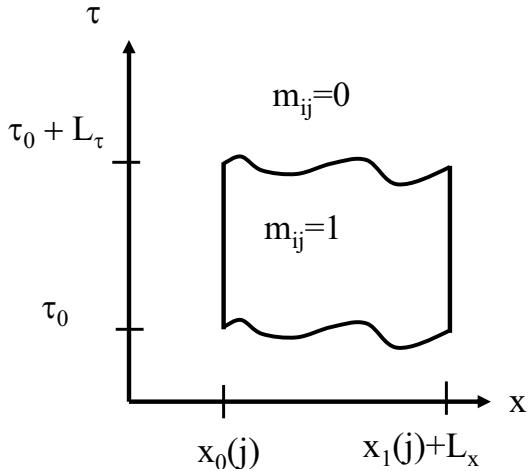}}
 \caption{An instanton starting from the ground state where $m_{i,j}=0$ towards a new
 minimum where $m_{ij}=1$ (see text).}
 \label{fig:instanton}
\end{figure}

In presence of the electric field $E$ any one of these ground
states (with $m_0$ fixed) become metastable since the phase $\phi$
wants to increase to gain energy from the field. We estimate the
tunnelling rate out of these metastable states if the electric
field $E$ is weak. They are given by $P \sim e^{- S^*_B/\hbar}$ to
exponential accuracy, where $S^*_B$ is the action of a bounce.
This is the instanton solution that corresponds to the minimal
action needed to go between the two minima and back
\cite{coleman_instanton_field,maki_instanton_periodic_long}. Such
an instanton has the shape of a bubble of typical size $L_x$ in
the space direction and $L_\tau$ in the time direction, and is
schematically represented in Fig.~\ref{fig:instanton}
If we denote $i,j$ the coordinate in space and time respectively, then
\begin{equation}
 \phi_{ij} = \phi^0_i + \delta\phi_{ij} = 2\pi (n_i + m_{ij}) + \zeta_i
\end{equation}
where $\delta\phi_{ij}$ is the deviation from the ground state. We
consider unit instantons with $m_{ij}=1$ inside the bubble, and
$m_{ij}=0$ outside. The region where $\phi_{ij}$ interpolates
between these two values is the wall which encircles the bubble,
which is very thin in the large $A^*=1$ limit considered here. It
is useful to recall that in the pure Sine-Gordon model (obtained
here taking all $\zeta_i=0$, $n_i^0=m_0$) the bounce instanton
solution (with zero friction coefficient) is a circle in the
$x,\tau$ plane, since the theory is Lorentz invariant.  Here, the
surface tension of the instanton walls is highly anisotropic. For
a ``time-like'' wall parallel to the $x$ axis, the surface tension
is the same as pure Sine-Gordon (with renormalized parameters)
since the disorder is time independant. The corresponding cost in
the action is $\sigma_\tau L_x$ where the line tension of such
walls is $\sigma_\tau = 2\pi/(K^* \alpha)$. For a ``space-like''
wall parallel to the $\tau$ axis the surface tension $\sigma_x(i)$
is a random variable. In first approximation the typical instanton
now has a rectangular shape, bounded in $x$ by two vertical
segments parallel to the $\tau$-axis at coordinates $x=i_0$ and
$x=i_0+L_x$, chosen as places where $\sigma_x(i)$ is small. The
rectangle is closed by two ``time-like'' segments at $\tau=\tau_0$
and $\tau=\tau_0 + L_\tau$.

Let us consider a segment of length $L_\tau$ of the wall parallel
to the $\tau$ direction between sites $i$ and $i+1$. The extra
action due to the presence of the instanton is:
\begin{eqnarray}
 \frac{\Delta S}{\hbar} = \frac{2 \pi}{K^* \alpha} \sum_j [ m_{i+1,j}^2 + 2
 m_{i+1,j} (n_{i+1}^0 - n_{i}^0 - f_i) ]
\end{eqnarray}
Thus, for a unit instanton $m_{i+1,j}=1$, the line tension
$\frac{\Delta S}{\hbar} = \sigma_x(i) L_\tau$, depends on
space position $i$:
\begin{eqnarray} \label{eq:tension}
\sigma_x(i) = \frac{4 \pi}{K^* \alpha} g_i
\end{eqnarray}
with $g_i=[f_i] - f_i + 1/2$. One easily sees that $g_i$ is
uniformly distributed on the interval $[0,1]$. In particular there
is a finite weight around $g_i=0$ which corresponds to ``weak
points'' in the construction of the ground state where one can
bifurcate from point $i$ up to the boundary at low energy cost to
a state where the phase is shifted by $2 \pi$ on the right of $i$
(or conversely $- 2 \pi$ on the left of $i$ for the wall on the
right). Although these states can be close in energy, the
tunnelling rate to them is zero. To obtain a non zero tunnelling
rate one must consider ``a kink'' i.e. tunnelling to a neighboring
state where the phase is shifted by $2 \pi$ between {\it two
walls}. This is the tunnelling process described by the above
instanton.

The total action cost of the above rectangular instanton is thus:
\begin{eqnarray}
 \frac{\Delta S}{\hbar} &=& (\sigma_x(i_0) + \sigma_x(i_0 + L_x))
 L_\tau + 2 \sigma_\tau L_x \nonumber \\
 & & - \frac{4\pi\epsilon L_x  L_\tau}{K ^* \alpha}
 \label{eq:cost}
\end{eqnarray}
Since the two smallest numbers in a set of $L_x/\alpha$ random
numbers are typically of order $\alpha/L_x$ one can estimate
$\sigma_x(i_0) + \sigma_x(i_0 + L_x) \sim \frac{4\pi}{K^* \alpha}
\frac{\alpha}{L_x}$. One then easily estimate the line tension
(\ref{eq:tension}) and by minimizing the action (\ref{eq:cost})
get the optimal size for the instanton (for small $\epsilon$)
\begin{eqnarray} \label{eq:saddle}
 L^{\text{opt}}_x = \sqrt{\alpha/\epsilon} &\quad,\quad&
 L^{\text{opt}}_\tau = 1/(2\epsilon)
\end{eqnarray}
This yields a decay rate:
\begin{eqnarray}
 P &\sim& e^{- \frac{4\pi}{K^* \alpha}
 \sqrt{\frac{\alpha}{\epsilon}}}
 = \text{exp}[- \frac{4\sqrt2\pi}{(K^*)^{3/2}}
 \sqrt{\frac{Q \Delta}{\rho_0 E \xi_{\text{loc}}}}]
\end{eqnarray}
where we have introduced a characteristic energy scale $\Delta =
\hbar u^* /\xi_{\text{loc}}$ associated with the localization
length. Note that $u^* /\xi_{\text{loc}}$ is the pinning frequency
\cite{fukuyama_pinning}. For a simple sine-Gordon theory the
dependence is $e^{- \Delta_M/(E \xi)}$ and $\Delta_M$ is the Mott
gap. This expression corresponds to Zener tunnelling across the
gap.

Although the above analysis is expected to give correctly the
electric field dependence, the precise prefactor in the
exponential might be modified by additional physical effects and
its precise determination, beyond the crude estimate given here,
is delicate. First strictly speaking, in order to reach a
stationary state some amount of dissipation should be introduced
in the model. This dissipation changes the cost of the time
variation of the phase and thus $\sigma_\tau$ but does not affect
$\sigma_x(i)$. It thus slightly changes the prefactor which could
in principle be studied as in \cite{hida_tunnel_sinegordon}. Next
since $\sigma_x(i_0) \neq \sigma_x(i_0 + L_x)$, the instanton has
a lozenge shape and the space like portion can improve its action
by taking advantage locally of favorable pins. That may slightly
renormalize downwards $\sigma_\tau$. Let us also point out that to
obtain the response of the system we have computed here a
\emph{typical} instanton, which can occur repeatedly in the volume
of the system. There are rarer events that correspond to faster
tunnelling. Let us divide the system in intervals of scale $R_x
\gg L^{\text{opt}}_x$ (up to the system size). Within each
interval there is typically one place to put two walls separated
by $L^{\text{opt}}_x$ and for which $(\sigma_x(i_0) + \sigma_x(i_0
+ L^{\text{opt}}_x) \sim 1/R_x$. Thus the ground state tunnels
(back and forth) with these states at a much faster rate. However
since these tunnelling events correspond to special places the
density of such atypical kinks being $O(L^{\text{opt}}_x/R_x)$
they cannot lead to a macroscopic current. The system thus stays
essentially blocked until the tunnelling events due to the typical
instantons can take place. It would however be interesting to see
whether such rare events could serve as nucleation center for
``quantum avalanches'', which could only increase the creep rate.
Although this clearly goes beyond the present study, the explicit
construction of the low lying states presented here may allow for
a precise study of this faster dynamics, at least numerically.

\begin{figure}
 \centerline{\includegraphics[width=\figwidth]{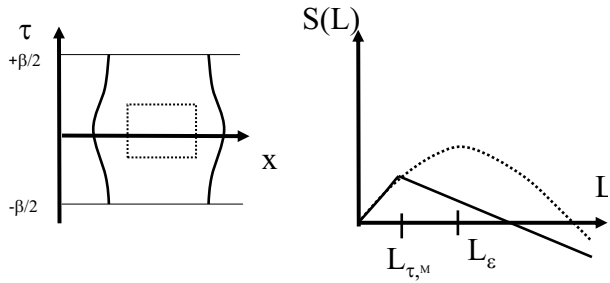}}
 \caption{Left: a bounce for large field (dashed line) and at very small fields (solid line). Since the
 bounce reaches the  finite size in imaginary time due to the finite temperature, it opens up. Right :
 The growth of the action at zero temperatures (dashed line) has a maximum which depends on the electric
 field. This is the barrier to pass to do the tunnelling. At finite temperature, the barrier decreases very rapidly
 when the bounce opens up (solid line). The barrier is thus determined essentially by the temperature.}
 \label{fig:openins}
\end{figure}

Let us now see how this quantum creep which corresponds to
tunnelling due to quantum effects is modified by the presence of a
non zero temperature. Because of the finite temperature the time
integration in imaginary time is limited to the finite value
$L_{\tau,M} = u^* \beta\hbar/(\xi_{\text{loc}}/\alpha)$ (because
of the rescaling). This means that the above analysis which was
done at $\beta=\infty$ remains valid as long as the size of the
bounce in the time direction is smaller than $L_{\tau,M}$. When
the size of the bounce reaches the boundary the instanton opens
(there are periodic boundary conditions in imaginary time) as
shown in Fig.~\ref{fig:openins}.
Because there is now no contribution coming from instantons
parallel to the space direction, it is easy to see from
(\ref{eq:cost}) that the action decreases linearly with the size
$L_x$ of the instanton. The tunnelling rate is thus fixed by the
maximum barrier, i.e. the value of the action when the bounce
reaches $L_{\tau,M}$. Because now the maximum barrier is not fixed
by the electric field any more, one has to consider both the
forward and backward jumps as for the standard TAFF argument
\cite{anderson_kim}. The net probability current is thus
proportional to
\begin{eqnarray}
 J &\propto & e^{-(S^*/\hbar - \epsilon \frac{4\pi L_{\tau,M}^2}{K^* \alpha})} -
              e^{-(S^*/\hbar + \epsilon \frac{4\pi L_{\tau,M}^2}{K^* \alpha})} \nonumber\\
  & \propto & e^{-\frac{S^*}{\hbar}} 2\sinh(\epsilon \frac{4\pi L_{\tau,M}^2}{K^* \alpha})
\end{eqnarray}
where $S^*$ is the action of the bounce as given by the saddle
point (\ref{eq:saddle}) when $L_\tau = L_{\tau,M}$. One thus
recovers below the crossover field $\epsilon =
\xi_{\text{loc}}/(2\beta \hbar \alpha u^*)$ a \emph{linear}
response, with a conductivity proportional to
\begin{equation}
 \sigma(T) \propto e^{-\frac{S^*}{\hbar}} =
 \text{exp}[- \frac{4\pi}{K^*}\sqrt{2\beta \Delta}]
\end{equation}
Quite remarkably the temperature dependence of the conductivity as
obtained by the present formula is identical to Mott's variable
range hopping \cite{mott_metal_insulator}, where the transition
between localized states close in energy is provided by external
source of inelastic scattering such as the electron phonon
interaction. The important difference between our result and the
standard VRH law is that here, inelastic processes are coming from
the electron-electron interaction itself (hidden in the existence
of the Luttinger liquid parameter $K$). Thus the \emph{prefactor}
in the exponential contains electronic energy scales. The VRH
formula contains normally the Debye temperature for phonons. Our
result thus lead to a quite different energy scale in the
exponential. Although our calculation is done in one dimension
only, it is most likely that in higher dimension as well one can
obtain similar formulas. Let us note that in one dimension the
above instanton picture is very similar physically to the VRH
picture, if one remembers that in a Luttinger liquid a kink in
$\phi$ is related to the presence of a charge though the formula
$\rho = - \nabla\phi/\pi$. Shifting the ground state by one unit
is equivalent to moving an electron.

\begin{acknowledgments}
TG would like to thank B. L. Altshuler for interesting
discussions.
\end{acknowledgments}


\begin{thebibliography}{26}
\expandafter\ifx\csname
natexlab\endcsname\relax\def\natexlab#1{#1}\fi
\expandafter\ifx\csname bibnamefont\endcsname\relax
  \def\bibnamefont#1{#1}\fi
\expandafter\ifx\csname bibfnamefont\endcsname\relax
  \def\bibfnamefont#1{#1}\fi
\expandafter\ifx\csname citenamefont\endcsname\relax
  \def\citenamefont#1{#1}\fi
\expandafter\ifx\csname url\endcsname\relax
  \def\url#1{\texttt{#1}}\fi
\expandafter\ifx\csname
urlprefix\endcsname\relax\def\urlprefix{URL }\fi
\providecommand{\bibinfo}[2]{#2}
\providecommand{\eprint}[2][]{\url{#2}}

\bibitem[{\citenamefont{Lemerle et~al.}(1998)\citenamefont{Lemerle, Ferr{\'e},
  Chappert, Mathet, Giamarchi, and {Le Doussal}}}]{lemerle_domainwall_creep}
\bibinfo{author}{\bibfnamefont{S.}~\bibnamefont{Lemerle et al.}},
  \bibinfo{journal}{Phys. Rev. Lett.}
  \textbf{\bibinfo{volume}{80}}, \bibinfo{pages}{849} (\bibinfo{year}{1998}).

\bibitem[{\citenamefont{Tybell et~al.}(2002)\citenamefont{Tybell, Paruch,
  Giamarchi, and Triscone}}]{tybell_ferro_creep}
\bibinfo{author}{\bibfnamefont{T.}~\bibnamefont{Tybell et al.}},
   \bibinfo{journal}{Phys. Rev. Lett.}
  \textbf{\bibinfo{volume}{89}}, \bibinfo{pages}{097601}
  (\bibinfo{year}{2002}).

\bibitem[{\citenamefont{Blatter et~al.}(1994)\citenamefont{Blatter, Feigel'man,
  Geshkenbein, Larkin, and Vinokur}}]{blatter_vortex_review}
\bibinfo{author}{\bibfnamefont{G.}~\bibnamefont{Blatter et al.}},
  \bibinfo{journal}{Rev. Mod. Phys.}
  \textbf{\bibinfo{volume}{66}}, \bibinfo{pages}{1125} (\bibinfo{year}{1994}).

\bibitem[{\citenamefont{Nattermann and
  Scheidl}(2000)}]{nattermann_vortex_review}
\bibinfo{author}{\bibfnamefont{T.}~\bibnamefont{Nattermann}} \bibnamefont{and}
  \bibinfo{author}{\bibfnamefont{S.}~\bibnamefont{Scheidl}},
  \bibinfo{journal}{Adv. Phys.} \textbf{\bibinfo{volume}{49}},
  \bibinfo{pages}{607} (\bibinfo{year}{2000}).

\bibitem[{\citenamefont{Gr{\"u}ner}(1988)}]{gruner_revue_cdw}
\bibinfo{author}{\bibfnamefont{G.}~\bibnamefont{Gr{\"u}ner}},
  \bibinfo{journal}{Rev. Mod. Phys.} \textbf{\bibinfo{volume}{60}},
  \bibinfo{pages}{1129} (\bibinfo{year}{1988}).

\bibitem[{\citenamefont{Andrei and {al.}}(1988)}]{andrei_wigner_2d}
\bibinfo{author}{\bibfnamefont{E.~Y.} \bibnamefont{Andrei}} \bibnamefont{and}
  \bibinfo{author}{\bibnamefont{{al.}}}, \bibinfo{journal}{Phys. Rev. Lett.}
  \textbf{\bibinfo{volume}{60}}, \bibinfo{pages}{2765} (\bibinfo{year}{1988}).

\bibitem[{\citenamefont{Willett and {\it et
  al.}}(1989)}]{willett_wigner_resistivity}
\bibinfo{author}{\bibfnamefont{R.~L.} \bibnamefont{Willett}} \bibnamefont{and}
  \bibinfo{author}{\bibnamefont{{\it et al.}}}, \bibinfo{journal}{Phys. Rev. B}
  \textbf{\bibinfo{volume}{38}}, \bibinfo{pages}{R7881} (\bibinfo{year}{1989}).

\bibitem[{\citenamefont{Giamarchi and Orignac}(2001)}]{giamarchi_quantum_revue}
\bibinfo{author}{\bibfnamefont{T.}~\bibnamefont{Giamarchi}} \bibnamefont{and}
  \bibinfo{author}{\bibfnamefont{E.}~\bibnamefont{Orignac}}, in
  \emph{\bibinfo{booktitle}{New Theoretical Approaches to Strongly Correlated
  Systems}}, edited by \bibinfo{editor}{\bibfnamefont{A.~M.}
  \bibnamefont{Tsvelik}} (\bibinfo{publisher}{Kluwer Academic Publishers},
  \bibinfo{address}{Dordrecht}, \bibinfo{year}{2001}).

\bibitem[{\citenamefont{Anderson and Kim}(1964)}]{anderson_kim}
\bibinfo{author}{\bibfnamefont{P.~W.} \bibnamefont{Anderson}} \bibnamefont{and}
  \bibinfo{author}{\bibfnamefont{Y.~B.} \bibnamefont{Kim}},
  \bibinfo{journal}{Rev. Mod. Phys.} \textbf{\bibinfo{volume}{36}},
  \bibinfo{pages}{39} (\bibinfo{year}{1964}).

\bibitem[{\citenamefont{Ioffe and Vinokur}(1987)}]{ioffe_creep}
\bibinfo{author}{\bibfnamefont{L.~B.} \bibnamefont{Ioffe}} \bibnamefont{and}
  \bibinfo{author}{\bibfnamefont{V.~M.} \bibnamefont{Vinokur}},
  \bibinfo{journal}{J. Phys. C} \textbf{\bibinfo{volume}{20}},
  \bibinfo{pages}{6149} (\bibinfo{year}{1987}).

\bibitem[{\citenamefont{Nattermann}(1987)}]{nattermann_rfield_rbond}
\bibinfo{author}{\bibfnamefont{T.}~\bibnamefont{Nattermann}},
  \bibinfo{journal}{Europhys. Lett.} \textbf{\bibinfo{volume}{4}},
  \bibinfo{pages}{1241} (\bibinfo{year}{1987}).

\bibitem[{\citenamefont{Maki}(1977)}]{maki_instanton_periodic_long}
\bibinfo{author}{\bibfnamefont{K.}~\bibnamefont{Maki}}, \bibinfo{journal}{Phys.
  Rev. B} \textbf{\bibinfo{volume}{18}}, \bibinfo{pages}{1641}
  (\bibinfo{year}{1977}).

\bibitem[{\citenamefont{Hida and Eckern}(1984)}]{hida_tunnel_sinegordon}
\bibinfo{author}{\bibfnamefont{K.}~\bibnamefont{Hida}} \bibnamefont{and}
  \bibinfo{author}{\bibfnamefont{U.}~\bibnamefont{Eckern}},
  \bibinfo{journal}{Phys. Rev. B} \textbf{\bibinfo{volume}{30}},
  \bibinfo{pages}{4096} (\bibinfo{year}{1984}).

\bibitem[{\citenamefont{Hida}(1984)}]{hida_tunnel_sinegordon_finiteT}
\bibinfo{author}{\bibfnamefont{K.}~\bibnamefont{Hida}}, \bibinfo{journal}{Z.
  Phys. B} \textbf{\bibinfo{volume}{61}}, \bibinfo{pages}{223}
  (\bibinfo{year}{1984}).

\bibitem[{\citenamefont{Shklovskii and
  Efros}(1981)}]{shklovskii_conductivity_coulomb}
\bibinfo{author}{\bibfnamefont{B.~I.} \bibnamefont{Shklovskii}}
  \bibnamefont{and} \bibinfo{author}{\bibfnamefont{A.~L.} \bibnamefont{Efros}},
  \bibinfo{journal}{Sov. Phys. JETP} \textbf{\bibinfo{volume}{54}},
  \bibinfo{pages}{218} (\bibinfo{year}{1981}).

\bibitem[{\citenamefont{Chauve et~al.}(2000)\citenamefont{Chauve, Giamarchi,
  and {Le Doussal}}}]{chauve_creep_long}
\bibinfo{author}{\bibfnamefont{P.}~\bibnamefont{Chauve}},
  \bibinfo{author}{\bibfnamefont{T.}~\bibnamefont{Giamarchi}},
  \bibnamefont{and} \bibinfo{author}{\bibfnamefont{P.}~\bibnamefont{{Le
  Doussal}}}, \bibinfo{journal}{Phys. Rev. B} \textbf{\bibinfo{volume}{62}},
  \bibinfo{pages}{6241} (\bibinfo{year}{2000}).

\bibitem[{\citenamefont{D.~A.~Gorokhov and
  Blatter}(2002)}]{gorokhov_quantum_creep}
\bibinfo{author}{\bibnamefont{D.~A.~Gorokhov}, \bibfnamefont{D.~S.~Fisher}}
  \bibnamefont{and} \bibinfo{author}{\bibfnamefont{G.}~\bibnamefont{Blatter}},
  \bibinfo{journal}{Phys. Rev. B}
  \textbf{\bibinfo{volume}{66}}, \bibinfo{pages}{214203} (\bibinfo{year}{2002}).

\bibitem[{\citenamefont{Balents and {Le Doussal}}(2002)}]{balents_frg_finiteT}
\bibinfo{author}{\bibfnamefont{L.}~\bibnamefont{Balents}} \bibnamefont{and}
  \bibinfo{author}{\bibfnamefont{P.}~\bibnamefont{{Le Doussal}}}
  (\bibinfo{year}{2002}), \bibinfo{note}{cond-mat/0205358}.

\bibitem[{\citenamefont{Larkin and Lee}(1978)}]{larkin_tunnelling_1d}
\bibinfo{author}{\bibfnamefont{A.~I.}~\bibnamefont{Larkin}} \bibnamefont{and}
  \bibinfo{author}{\bibfnamefont{P.~A.} \bibnamefont{Lee}},
  \bibinfo{journal}{Phys. Rev. B} \textbf{\bibinfo{volume}{17}},
  \bibinfo{pages}{1596} (\bibinfo{year}{1978}).

\bibitem[{\citenamefont{Mott}(1990)}]{mott_metal_insulator}
\bibinfo{author}{\bibfnamefont{N.~F.} \bibnamefont{Mott}},
  \emph{\bibinfo{title}{Metal-Insulator Transitions}}
  (\bibinfo{publisher}{Taylor and Francis}, \bibinfo{address}{London},
  \bibinfo{year}{1990}).

\bibitem[{\citenamefont{Fukuyama and Lee}(1978)}]{fukuyama_pinning}
\bibinfo{author}{\bibfnamefont{H.}~\bibnamefont{Fukuyama}} \bibnamefont{and}
  \bibinfo{author}{\bibfnamefont{P.~A.} \bibnamefont{Lee}},
  \bibinfo{journal}{Phys. Rev. B} \textbf{\bibinfo{volume}{17}},
  \bibinfo{pages}{535} (\bibinfo{year}{1978}).

\bibitem[{\citenamefont{Giamarchi and Schulz}(1988)}]{giamarchi_loc}
\bibinfo{author}{\bibfnamefont{T.}~\bibnamefont{Giamarchi}} \bibnamefont{and}
  \bibinfo{author}{\bibfnamefont{H.~J.} \bibnamefont{Schulz}},
  \bibinfo{journal}{Phys. Rev. B} \textbf{\bibinfo{volume}{37}},
  \bibinfo{pages}{325} (\bibinfo{year}{1988}).

\bibitem[{\citenamefont{Glatz and Nattermann}(2002)}]{glatz_cdw_finitetemp}
\bibinfo{author}{\bibfnamefont{A.}~\bibnamefont{Glatz}} \bibnamefont{and}
  \bibinfo{author}{\bibfnamefont{T.}~\bibnamefont{Nattermann}},
  \bibinfo{journal}{Phys. Rev. Lett.} \textbf{\bibinfo{volume}{88}},
  \bibinfo{pages}{256401} (\bibinfo{year}{2002}).

\bibitem[{\citenamefont{Glatz}(2001)}]{glatz_diplom}
\bibinfo{author}{\bibfnamefont{A.}~\bibnamefont{Glatz}} (\bibinfo{year}{2001}),
  \bibinfo{note}{diploma thesis, Cologne 2001, cond-mat/0302133.}

\bibitem[{\citenamefont{Giamarchi and {Le
  Doussal}}(1996)}]{giamarchi_columnar_variat}
\bibinfo{author}{\bibfnamefont{T.}~\bibnamefont{Giamarchi}} \bibnamefont{and}
  \bibinfo{author}{\bibfnamefont{P.}~\bibnamefont{{Le Doussal}}},
  \bibinfo{journal}{Phys. Rev. B} \textbf{\bibinfo{volume}{53}},
  \bibinfo{pages}{15206} (\bibinfo{year}{1996}).

\bibitem[{\citenamefont{Coleman}(1977)}]{coleman_instanton_field}
\bibinfo{author}{\bibfnamefont{S.}~\bibnamefont{Coleman}},
  \bibinfo{journal}{Phys. Rev. D} \textbf{\bibinfo{volume}{15}},
  \bibinfo{pages}{29292} (\bibinfo{year}{1977}).

\end{thebibliography}
\end{document}